\documentclass{PoS}

\usepackage{graphicx}
\usepackage{amssymb}
\usepackage{amsmath}
\usepackage{bbold}
\usepackage{url}
\usepackage{natbib}
\bibpunct{(}{)}{;}{a}{}{,}

\newcommand{\twofig}[4]{%
  \begin{figure*}%
    \centerline{\resizebox{\hsize}{!}{\includegraphics*{#1} \,%
        \includegraphics*{#2}}}%
    \caption{#4}\label{#3}%
  \end{figure*}%
}

\newcommand{\threefig}[5]{%
  \begin{figure*}%
    \centerline{\resizebox{0.56\hsize}{!}{\includegraphics*{#1}}}
    \centerline{\resizebox{0.56\hsize}{!}{\includegraphics*{#2}}}
    \centerline{\resizebox{0.56\hsize}{!}{\includegraphics*{#3}}}
    \caption{#5}\label{#4}%
  \end{figure*}%
}

\newcommand{\sect}[1]{Sect.~\ref{#1}}

\newcommand{\fig}[1]{Fig.~\ref{#1}}

\newcommand{\eq}[1]{Eq.~(\ref{#1})}

\renewcommand{\S}{\mathbf{S}}
\newcommand{\N}{\mathbf{N}}
\newcommand{\Nbar}{\mathbf{\bar{N}}}
\newcommand{\T}{\mathbf{T}}
\newcommand{\lmax}{\ell_{\mathrm{max}}}

\newcommand{\order}[1]{${{\cal O}\! \left( #1 \right)}$}
\newcommand{\beq}{\begin{equation}}
\newcommand{\eeq}{\end{equation}}
\newcommand{\wmap}{\emph{WMAP}}

\def\aap{A\&A}
\def\apj{ApJ}
\def\apjl{ApJ}
\def\apjs{ApJS}

\def\prd{Phys.~Rev.~D}

\title{Fast Wiener filtering of CMB maps}

\ShortTitle{Fast Wiener filtering of CMB maps}

\author{Franz Elsner\\
  Institut d'Astrophysique de Paris, UMR 7095, CNRS -
  Universit\'e Pierre et Marie Curie (Univ Paris 06), 98 bis blvd
  Arago, 75014 Paris, France\\
  E-Mail: \email{elsner@iap.fr}}

\author{\speaker{Benjamin D. Wandelt}\\
  Institut d'Astrophysique de Paris, UMR 7095, CNRS -
  Universit\'e Pierre et Marie Curie (Univ Paris 06), 98 bis blvd
  Arago, 75014 Paris, France\\
  Departments of Physics and Astronomy, University of Illinois at
  Urbana-Champaign, Urbana, IL 61801, USA\\
  E-Mail: \email{benwandelt@gmail.com}}

\abstract{We present the application of a new method to compute the
  Wiener filter solution of large and complex data sets. Contrary to
  the iterative solvers usually employed in signal processing, our
  algorithm does not require the use of preconditioners to be
  computationally efficient. The new scheme is conceptually very
  simple and therefore easy to implement, numerically absolutely
  stable, and guaranteed to converge. We introduce a messenger field
  to mediate between the different preferred bases in which signal and
  noise properties can be specified most conveniently, and rephrase
  the signal reconstruction problem in terms of this auxiliary
  variable. We demonstrate the capabilities of the algorithm by
  applying it to cosmic microwave background (CMB) radiation data
  obtained by the \wmap\ satellite.}

\FullConference{Big Bang, Big Data, Big Computers,\\
		September 19-21, 2012\\
		Laboratoire Astroparticule et Cosmologie, 10 rue
                A. Domon et L. Duquet, 75205 Paris 13, France}

\begin{document}

\section{Introduction}
\label{sec:intro}

Among the linear filters that make use of statistical information
about the data, the generalized Wiener filter stands out as the
maximum a posteriori solution for the case that signal and noise both
follow a Gaussian distribution \citep{1949wiener}. Although Wiener
filtering is of fundamental importance, only recently efficient
algorithms have been developed for the relaxed assumption that the
signal covariance is not known in advance
\citep[e.g.,][]{2004PhRvD..70h3511W, 2011PhRvD..83j5014E}.

Assuming a linear model for the observed data $d$ as a combination of
signal $s$ and noise $n$,
\beq
\label{eq:def_data}
d = s + n\, ,
\eeq
the Wiener filter $s_{\mathrm{WF}}$ is defined as the solution of the
equation
\beq
\label{eq:def_wiener}
(\S^{-1} + \N^{-1}) \, s_{\mathrm{WF}} = \N^{-1} d \, .
\eeq
It plays an important role in signal processing. Taking the example of
cosmic microwave background (CMB) radiation data, it is calculated
during optimal power spectrum estimation
\citep[e.g.,][]{1997PhRvD..55.5895T, 1998PhRvD..57.2117B,
  1999ApJ...510..551O, 2012A&A...540L...6E}, likelihood analysis
\citep[e.g.,][]{2007ApJS..170..288H, 2009ApJS..180..306D,
  2012A&A...542A..60E}, mapmaking \citep[e.g.,][]{1997ApJ...474L..77T,
  1997ApJ...480L..87T}, and lensing reconstructions
\citep[e.g.,][]{2004PhRvD..70j3501H, 2007PhRvD..76d3510S}, etc.

To evaluate \eq{eq:def_wiener} in practice can turn out very
challenging for the large and complex data sets obtained by state of
the art experiments. Problems occur as the size of $\S$ and $\N$
increase with the second power of the number of data samples,
rendering the storage and processing of dense systems
impractical. Fortunately, it is often feasible to specify sets of
bases where signal and noise covariance become sparse. However, these
bases are generally incompatible, i.e., it is not possible to
represent them in a single basis as sparse systems such that
\eq{eq:def_wiener} can be solved trivially. Numerical algorithms for
the exact solution of the Wiener filter equation are therefore often
complex, e.g., involving conjugate gradient solvers with multigrid
preconditioners \citep{2007PhRvD..76d3510S}. Finding fast
preconditioners that work is a highly non-trivial art, especially
since the matrices of interest are often extremely ill-conditioned
($\lambda_{\mathrm{max}} / \lambda_{\mathrm{min}} \gtrsim 10^{7}$).

In \sect{sec:method}, we briefly describe a new iterative algorithm
for the solution of the exact Wiener filter equation. We illustrate
the approach by applying it to \wmap\emph{7} data in
\sect{sec:wmap}. Summarizing the main aspects of our findings, we then
discuss possible extensions and areas of applications
(\sect{sec:summary}). Details on the mathematical aspects of the
method can be found in \cite{2013A&A...549A.111E}.

\section{Method}
\label{sec:method}

To efficiently compute the Wiener filter, we first introduce a
messenger field $t$ with covariance $\T$. The covariance properties of
this auxiliary variable are very simple: $\T$ is proportional to the
identity matrix, a property that is preserved under \emph{any}
orthogonal basis change. Therefore, while it may not be possible to
directly apply expressions like $(\S + \N)^{-1}$ to a data vector, we
can always do so with combinations as $(\S + \T)^{-1}$ and $(\N +
\T)^{-1}$, no matter what basis is chosen to render $\S$ and $\N$
sparse.

To benefit from this possibility, we now specify equations for the
signal reconstruction $s$ and the auxiliary field $t$,
\begin{align}
 \label{eq:def_algorithm1}
    \left( \Nbar^{-1} + (\lambda \T)^{-1} \right) \,
    t & = \Nbar^{-1} \, d + (\lambda \T)^{-1} \, s
    \\
 \label{eq:def_algorithm2}
    \left( \S^{-1} + (\lambda \T)^{-1} \right)
    \, s & = (\lambda \T)^{-1} \, t \, ,
\end{align}
where we defined $\Nbar \equiv \N - \T$ and introduced one additional
scalar parameter $\lambda$, which we will use to accelerate
convergence. We specify the covariance matrix of the auxiliary field
according to $\mathbf{T} = \min(\mathrm{diag}(\mathbf{N})) \cdot
\mathbb{1} $. In the limit $\lambda = 1$, the system of equations
defined above reduces to the Wiener filter equation.

An outline of the algorithm can be given as follows. Initially, the
vectors $s$ and $t$ are set to zero. We solve \eq{eq:def_algorithm1}
for the auxiliary field $t$ in the basis defined by the noise
covariance matrix. Next, we change the basis to, e.g., Fourier space,
where $\S$ can be represented easily. Then, we solve for $s$ using
\eq{eq:def_algorithm2} given the latest realization of $t$, and
finally transform the result back to the original basis. It can be
shown that the signal reconstruction $s$ converges exponentially and
unconditionally to the Wiener filter solution $s_{\mathrm{WF}}$
\citep{2013A&A...549A.111E}. Based on a comparison to more standard
conjugate gradient solvers, we find the final map to be accurate to
about 1 part in $10^5$, depending on the adopted stopping criterion.

\section{\wmap\ analysis}
\label{sec:wmap}

We now demonstrate the application of the algorithm to the V-band data
of the Wilkinson microwave anisotropy probe
\citep[\wmap\emph{7\footnote{Available from
      \url{http://lambda.gsfc.nasa.gov}}},][]{2011ApJS..192...14J}. We
adopted the official \wmap\ extended temperature and polarization
masks, reducing the sky fraction to around 70~\%.

For the best performance in terms of computing time, we changed
$\lambda$ in the following way: To initialize the algorithm, we chose
it according to the ratio $T / C^\mathrm{TT}_{\ell_{\mathrm{iter}}}$
at $\ell_{\mathrm{iter}} = 20$. As shown in \fig{fig:kernel}, a high
value of $\lambda$ corresponds to a large convolution kernel in
\eq{eq:def_algorithm2}. Accordingly, small scales are smoothed out and
information is more efficiently propagated into the masked regions. To
monitor convergence of the algorithm, we used the $\chi^2 =
s^{\dagger} \S^{-1} s + (d - s)^{\dagger} \N^{-1} (d - s)$ of the
current solution $s$. After it has reached a plateau, we decreased
$\lambda$ to the ratio at $\ell^{\mathrm{new}}_{\mathrm{iter}} = 2 \,
\ell^{\mathrm{old}}_{\mathrm{iter}}$. The evolution of $\lambda$,
together with the current value of $\chi^2$ at every iteration, is
depicted in \fig{fig:chi2_lambda}. We stopped the algorithm as soon as
the quality of the signal reconstruction improved only marginally
between iterations, $\Delta \chi^2 < 10^{-4} \, \sigma_{\chi^2}$ at
$\lambda = 1$, and obtain a final goodness of fit of $\chi^2 /
n_{\mathrm{d.o.f.}} = 0.996$.

Since we can attribute a specific length scale to a given value of
$\lambda$ (the current value of $\ell_{\mathrm{iter}}$), and
fluctuations on smaller scales are suppressed owing to the width of the
convolution kernel, we can restrict the maximum multipole moment of
the computationally expensive spherical harmonic transforms to
$\lmax^{\mathrm{SHT}} \approx \ell_{\mathrm{iter}} + 100$. As a
result, the overall computing time for the spherical harmonic
transforms can be reduced by a large margin. We note that for the most
expensive final iterations, where $\lmax^{\mathrm{SHT}} \approx \lmax$
and more than half of the computing time is spent, the algorithm is an
excellent candidate for further acceleration by means of graphics
processing units \citep{2011A&A...532A..35E}.

The buildup of the \wmap\ temperature Wiener filtered map is shown in
\fig{fig:wf_temperature}. Characteristic for Wiener filter solutions,
large scale fluctuations extend well inside masked regions. In
\fig{fig:power_spectra}, we plot the power spectra of the
reconstruction for temperature and polarization, and demonstrate that
the Wiener filter maps can be augmented to constrained realizations,
with the correct signal variances.

\twofig{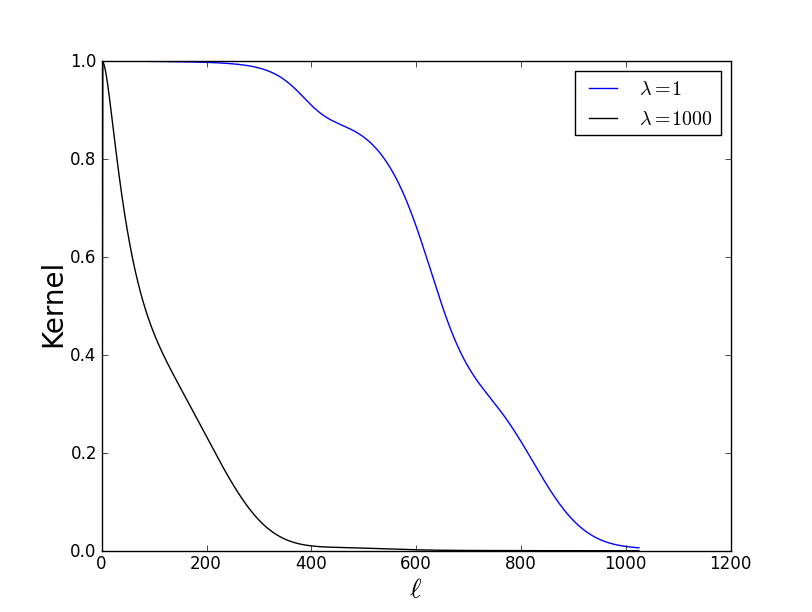}{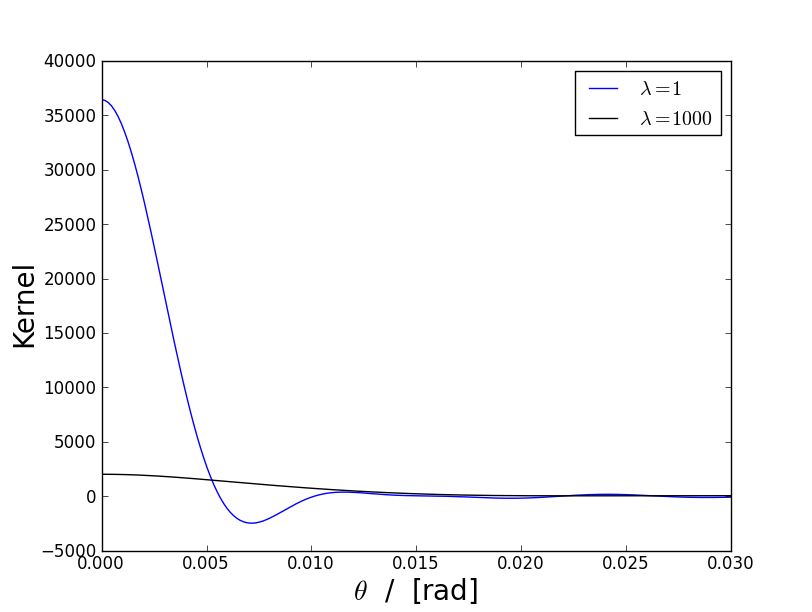}{fig:kernel}%
{Convolution kernel. For $\lambda = 1$ (\emph{blue lines}) and
  $\lambda = 1000$ (\emph{black lines}), we show the shape of the
  convolution kernel used to solve for the signal given the current
  reconstruction of the messenger field in harmonic space (\emph{left
    panel}) and real space (\emph{right panel}).}

\twofig{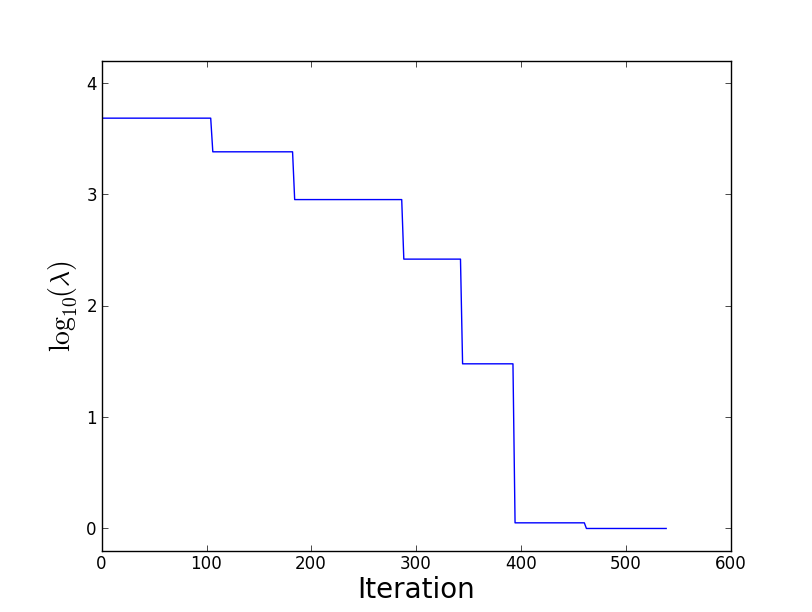}{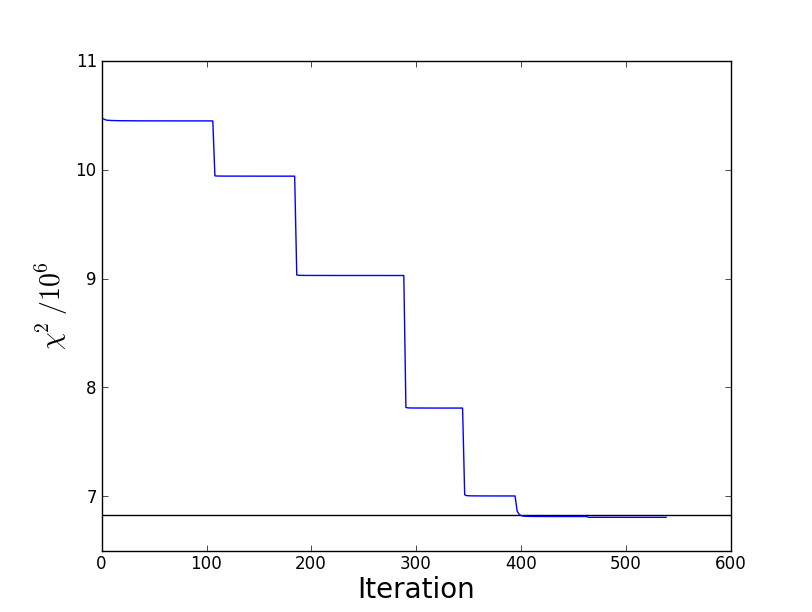}{fig:chi2_lambda}
{Convergence of the algorithm. \emph{Left panel:} Starting from
  \order{10^4}, we reduce $\lambda$ step by step to unity. \emph{Right
    panel:} For a given $\lambda$, the $\chi^2$ of the solution first
  drops quickly and then reaches a plateau (\emph{blue line}). The
  expectation value of $\chi^2$ for the final solution, given by the
  number of degrees of freedom, is also indicated (\emph{black
    line}).}

\threefig{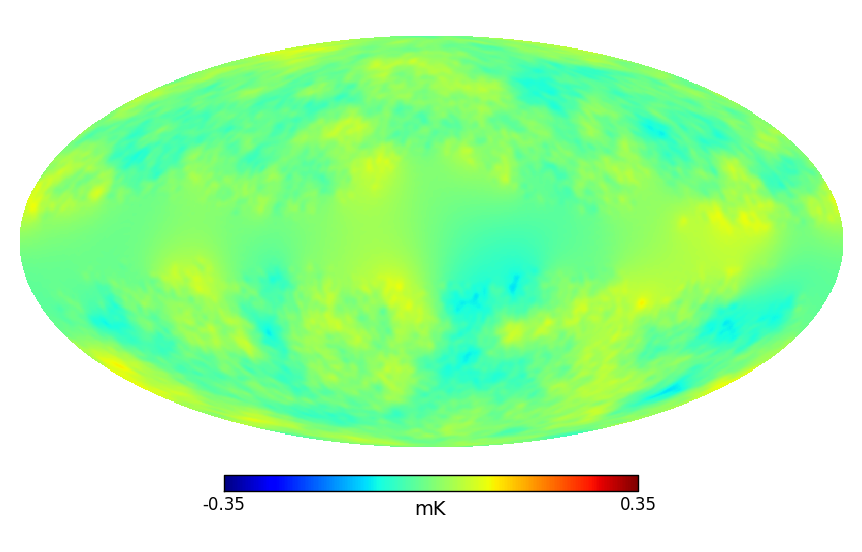}{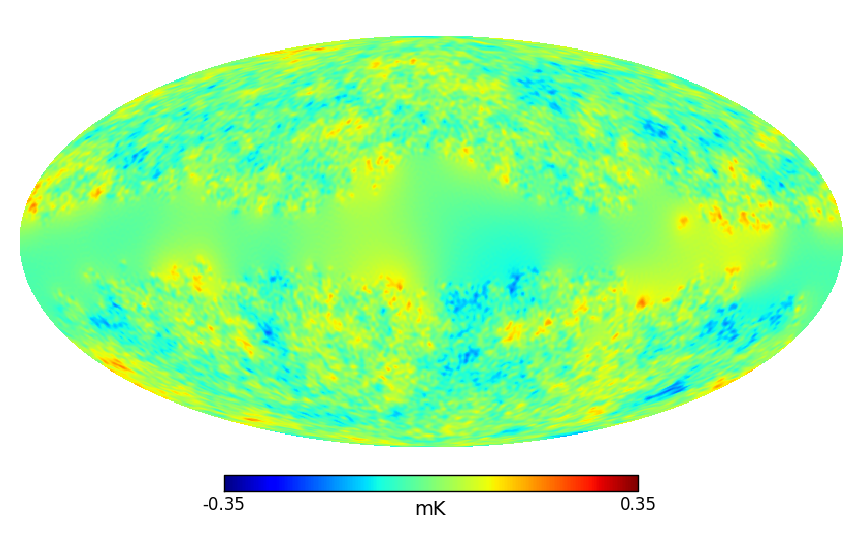}%
{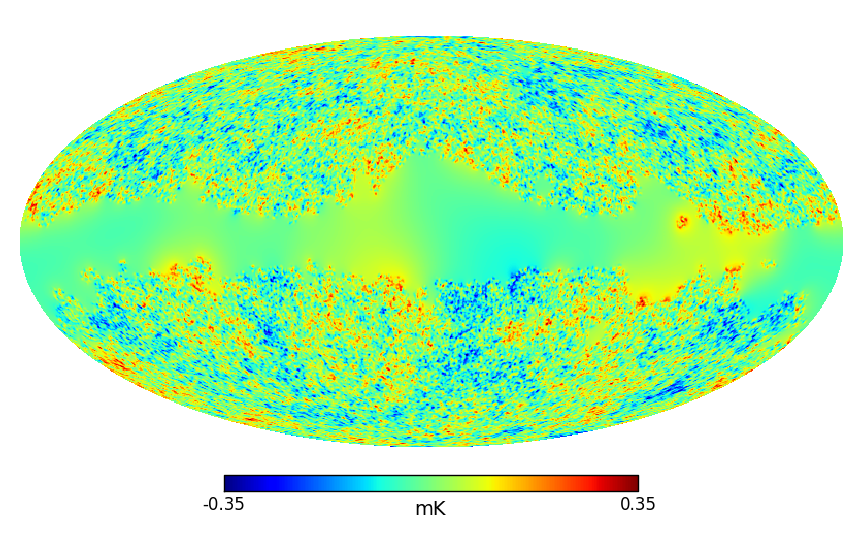}{fig:wf_temperature}%
{Buildup of the temperature reconstruction. Snapshots of the algorithm
  after 50 iterations (\emph{upper panel}), 250 iterations
  (\emph{middle panel}), and of the final solution (\emph{lower
    panel}) show the convergence of the reconstructed signal from the
  largest to the smallest scales.}

\threefig{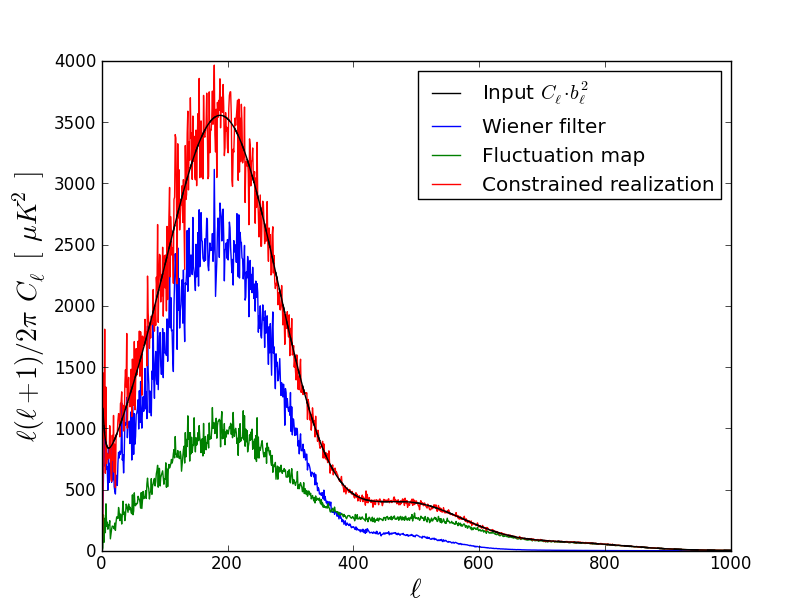}{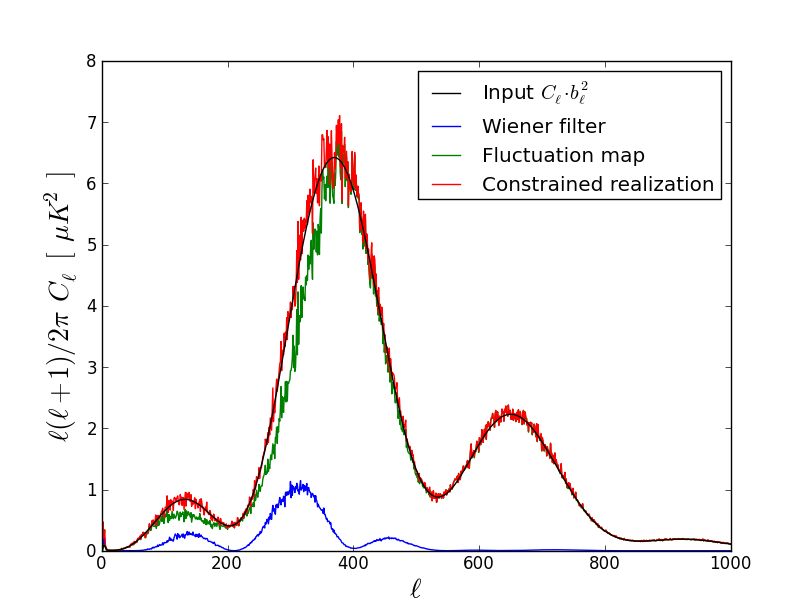}%
{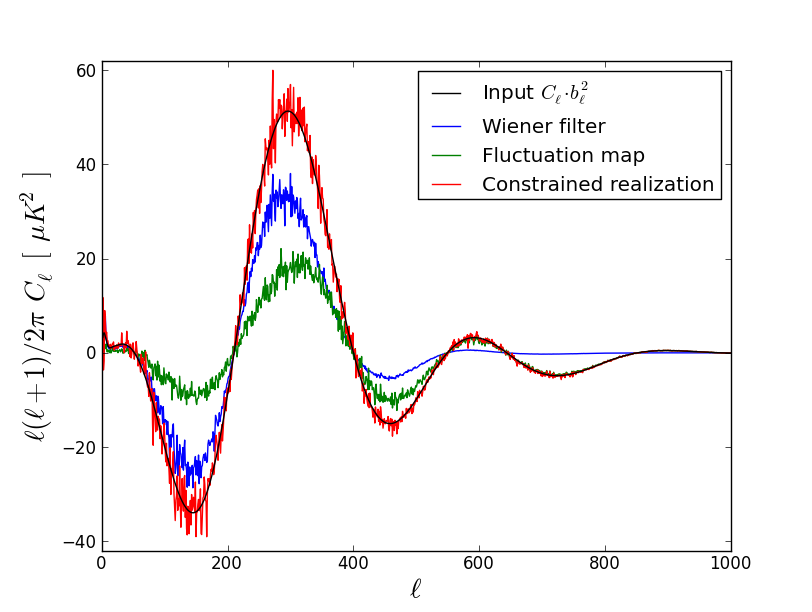}{fig:power_spectra}%
{Power spectra of the reconstruction. In this series of plots of the
  temperature (TT, \emph{upper panel}), polarization (EE, \emph{middle
    panel}), and cross power spectra (TE, \emph{lower panel}), we show
  that a constrained realization obtained with this algorithm
  (\emph{red line}), consisting of the the sum of the Wiener filter
  solution (\emph{blue line}) and a fluctuation map (\emph{green
    line}), is unbiased compared to the fiducial power spectrum
  multiplied by the beam function (\emph{black line}).}

\section{Conclusion}
\label{sec:summary}

We have summarized a new method to efficiently calculate the Wiener
filter solution of general data sets. As a sample application, we
analyzed \wmap\ temperature and polarization maps.

The algorithm is not only simple to implement, but also robust. Even
after including polarization data in the analysis, it maintains good
performance, despite of the increase in the condition number of the
covariance matrices. We consider this fact to be of particular
importance as conjugate gradient solvers are likely to perform
noticeably worse in this setup.

We finally note that the algorithm is also very flexible, and many
possible extensions are still to be explored. For some problems, for
example, it may be beneficial to solve for the noise vector instead of
the signal, which is calculated indirectly from the result. The
auxiliary field then becomes associated with the signal reconstruction
instead of the noise. In certain situations, it may also prove useful
to include more than one messenger field, for example if multiple
observations (e.g., from different detectors) should be combined in a
joint analysis. It also remains to be explored to what extent our
method could be combined with conventional iterative schemes, for
example to act as a smoother in a multigrid scheme to speed up
convergence further.

\end{document}